\documentclass[showpacs,superscriptaddress, preprintnumbers,amsmath,amssymb]{revtex4}

\usepackage{graphicx}
\usepackage{dcolumn}
\usepackage{bm}

\begin{document}

\preprint{APS/123-QED}

\title{Cluster structure of EU-15 countries derived from the correlation matrix analysis of macroeconomic index fluctuations}

\author{M. Gligor}
\affiliation{SUPRATECS, Universit\'e de Li\`ege, B5 Sart-Tilman, B-4000 Li\`ege, Belgium
}
\affiliation{National College 'Roman Voda', Roman-5550, Neamt, Romania}%
\author{M. Ausloos}
\affiliation{SUPRATECS, Universit\'e de Li\`ege, B5 Sart-Tilman, B-4000 Li\`ege, Belgium
}

\date{10/07/2006}

\begin{abstract}
The statistical distances between countries, calculated for various moving average time windows, are mapped into the ultrametric subdominant space as in classical Minimal Spanning Tree methods. The Moving Average Minimal Length Path (MAMLP) algorithm allows a decoupling of fluctuations with respect to the mass center of the system from the movement of the mass center itself. A Hamiltonian representation given by a factor graph is used and plays the role of cost function. The present analysis pertains to 11 macroeconomic (ME) indicators, namely the GDP ($x_1$), Final Consumption Expenditure ($x_2$), Gross Capital Formation ($x_3$), Net Exports ($x_4$), Consumer Price Index ($y_1$), Rates of Interest of the Central Banks ($y_2$), Labour Force ($z_1$), Unemployment ($z_2$), GDP/hour worked ($z_3$), GDP/capita ($w_1$) and Gini coefficient ($w_2$). The target group of countries is composed of 15 EU countries, data taken between 1995 and 2004. By two different methods (the Bipartite Factor Graph Analysis and the Correlation Matrix Eigensystem Analysis) it is found that the strongly correlated countries with respect to the macroeconomic indicators fluctuations can be partitioned into stable clusters.

\end{abstract}


\maketitle

\section{Introduction}

Modeling the dependences between the macroeconomic (ME) variables has to take into account circumstances that differ substantially from those encountered in the natural sciences. First, experimentation is usually not feasible and is replaced by survey research, implying that the explanatory variables cannot be manipulated and fixed by the researcher. Second, the number of possible explanatory variables is often quite large, unlike the small number of carefully chosen treatment variables frequently found in the natural sciences. Third, the ME time series are short and noisy. Most data have a yearly frequency. When social time series have been produced for a very long period, there is usually strong evidence against stationarity.
\par Some macroeconomic (ME) indicators are monthly and/or quarterly registered, increasing in this way the number of available data points, but some additional noise is naturally enclosed in the time series so generated (seasonal fluctuations, external and internal short range shocks, etc). This seems to be a solid argument for the fact that the main data sources, at least the ones freely available on the web, tend only to keep the annual averages/rates of growth of the ME indicators.
\par Let us consider, for example, a time interval of one hundred years, which is mapped onto a graphical plot of 100 data points. From the statistical physics viewpoint, 100 is a quite small number of data points, surely too small for speaking about the so called \textit{thermodynamic limit}. On the other hand, from a socio-economic point of view, we can justifiably wonder if a growth, say, of 2\% of any ME indicator has at the present time the same meaning as it had one century ago. One must take into account that during that time, the social, politic and economic environment was drastically changed. Moreover the methodology of data collecting and processing is today different from what it was two generations ago. Indeed, the economic world is created by people and is substantially changing from a generation to another one (sometimes also during one and the same generation). Thus, this way of statistical data aggregation turns to be controversial.
\par Several papers \cite{canning, amaral} investigated the statistical patterns in GDP annual rates of growth by aggregating (in a "horizontal" way) the data from all countries for which statistical data were reported. Even if all data are supposed to be reliable, and even if the relative rates of growth are investigated (to diminish the actual large difference influences), this way of aggregation, as well as the previous one, supposes a priori a certain degree of homogeneity across countries. A certain GDP rate of growth in an underdeveloped country is certainly based on factors that differ substantially from the ones that generate the same rate of growth in a developed country. Both theoretical and empirical investigations \cite{durlauf, hill} reported the evidence of the country partitioning in clusters after their common patterns of evolution. For such subsystems only, the data might be meaningfully aggregated. In the present paper we demonstrate \textit{the clustering emergence in the relatively stable and homogeneous system} composed of the 15 EU countries for data taken between 1994 and 2004, starting from the annual rates of growth of 11 ME indicators, namely the GDP ($x_1$), Final Consumption Expenditure ($x_2$), Gross Capital Formation ($x_3$), Net Exports ($x_4$), Consumer Price Index ($y_1$), Rates of Interest of the Central Banks ($y_2$), Labour Force ($z_1$), Unemployment ($z_2$), GDP/hour worked ($z_3$), GDP/capita ($w_1$) and Gini coefficient ($w_2$).
\par One has to stress here that the problem of studying the patterns of growth across countries is actually a subject of great attention to economists \cite{hill, chen}. An important reason for the increasing interest in this problem is that persistent disparities in aggregate growth rates across countries have, over time, led to large differences in welfare. On the other hand, the intellectual payoffs are high: various statistical tools might be considerably enriched and extended by applying them to the non-stationary, short and noisy macroeconomic time series.
\par In the present paper we focus on two recent lines of research, of growing interest in physics, which can bring important contributions to ME time series analysis. On one hand, the recent developments in nonequilibrium networks \cite{net}; on the other hand, the random matrix theory (RMT), initially developed in nuclear physics, also successfully used in the study of canonical correlations between stock changes and portfolio optimization problem \cite{risk}. The way in which these methods are adapted to the macroeconomic time series analysis is described in the next section.
\par The Minimal Spanning Tree (MST) is one of the most usual methods in cluster analysis, and has been largely used so far both by physicists \cite{dimatteo} and economists \cite{hill}. Nonetheless, both sides \cite{hill, risk} noted some lack of univocity due to choosing the MST root. Moreover, the MST structure proves to be not stable when a constant size time window is moved over the considered time span. The solution briefly presented in Section 3, namely the Moving Average Minimal Length Path (MAMLP) method comes as a development of some previous methods where some arbitrariness in the root of the tree was underlined considering that an a priori more common root, like the sum of the data, called the "All" country, from which to let the tree grow was permitting a better comparison \cite{misk}.
\par The target group of countries is composed of 15 EU countries, data taken between 1994 and 2004. The main sources used for all the above indicators annual rates is the World Bank database \cite{wb} and the OECD database \cite{oecd}. We abbreviate the countries according to the Roots Web Surname List (RSL) which uses 3 letters standardized abbreviations to designate countries and other regional locations (http://helpdesk.rootsweb.com/codes/). Inside the tables, for spacing reasons we use the countries two letters abbreviation (http://www.iso.org).
\par The remainder of the paper is organized as follows: in Section 2 the theoretical and methodological tools from the network analysis and matrix theory which we try to adapt to the considered time series are briefly described. The results are largely presented and discussed in Section 3. Some concluding remarks are done in Section 4.
   
\section{Theoretical and methodological framework}

As mentioned in Sect. 1, MST cannot be built in a unique way, whence this becomes a problem when we try to construct a cluster hierarchy for each position of a moving time window. The hierarchical structure proved to be not robust against fluctuations induced by a moving time window.
      In the MAMLP method described here below we propose to construct the hierarchy starting from a virtual 'average' agent. The method is developed in the following steps:
(i)    An "AVERAGE' agent (AV) is virtually included into the system; the statistical distance matrix is constructed, and thereafter, the elements are set into increasing order (i.e. the decreasing order of correlations);
(ii)   The hierarchy is constructed, connecting each agent by its minimal length path (MLP) to AV. Its minimal distance to AV is associated to each agent.
(iii) The procedure is repeated by moving a given and constant time window (in this case a 5 years time window size) over the investigated time span (in the present analysis: 1994-2004). The agents are sorted through their movement inside the hierarchy. Therefore, a new correlation matrix between country distances to their own mean is constructed. The matrix elements are defined as:

\begin{equation}
\hat{C}_{i,j}(t)= {\frac{<\hat{d}_{i}(t)  \hat{d}_{j}(t)>-<\hat{d}_{i}(t)>  <\hat{d}_{j}(t)>}{\sqrt{<(\hat{d}_{i}(t))^{2}-<\hat{d}_{i}(t)>^{2}>  <(\hat{d}_{j}(t))^{2}-<\hat{d}_{j}(t)>^{2}>}}}
\end{equation}

\noindent where $\hat{d}_{i}(t)$ is the i-country minimal length path (MPL) distance to the AVERAGE. For simplicity, the explicit dependencies on the time window size $T$ are not included in Eq. (1).

\begin{table}

\caption{MPL distances to AVERAGE. The moving time window size is 5 years for data taken from 1994 to 2004.}
\bigskip
\begin{footnotesize}
\begin{center}
\begin{tabular}{c c c c c c c c c c c c c c c c}

 & AU & BE & DE	& DK	& ES	& FI	& FR	& UK	& GR	& IE	& IT	& LU	& NL	& PT	& SE
\\\hline 
\\94-98 & .67 & .86 &	.86 &	.86 &	.40 &	.40 &	.67 &	.86 &	.40 &	.86 &	.86 &	.40 &	.40 &	.86 &	.86\\
\\95-99 &  .60 &	.65 &	.52 &	.71 &	.21 &	.77 &	.45 &	.77 &	.37 &	.65 &	.90 &	.37 &	.23 &	.83 &	.52\\
\\ 96-00 &	.58 &	.32 &	.46 &	.61 &	.34 &	.81 &	.46 &	.32 &	.32 &	.53 &	.32 &	.20 &	.60 &	.60 &	.46 \\
\\ 97-01 &	.48 &	.30 &	.48 &	.30 &	.28 &	.42 &	.48 &	.44 &	.68 &	.38 &	.68 &	.14 &	.28 &	.28 &	.48 \\
\\ 98-02 &	.43 &	.26 &	.19 &	.19 &	.21 &	.43 &	.19 &	.19 &	1.04 &	.29 &	.44 &	.12 &	.21 &	.21  &	.29 \\
\\ 99-03 &	.25 &	.23 &	.19 &	.19 &	.29 &	.26 &	.19 &	.37 &	1.15 &	.26 &	.37 &	.23 &	.19 & 	.19 &	.28 \\
\\ 00-04 &	.27 &	.27 &	.17 &	.26 &	.28 &	.27 &	.21 &	.27 &	.53 &	.50 &	.28 &	.27 &	.21 &	.21&	.27 \\
\\\hline

\end{tabular}
\end{center}
\end{footnotesize}
\end{table}

\begin{table}

\caption{The correlation matrix of EU-15 country movements inside the hierarchy. Indicator: GDP. The moving time window size is 5 years for data taken from 1994 to 2004.}
\bigskip
\begin{footnotesize}
\begin{center}
\begin{tabular}{c c c c c c c c c c c c c c c c}

 & AU & BE & DE	& DK	& ES	& FI	& FR	& UK	& GR	& IE	& IT	& LU	& NL	& PT	& SE
\\\hline 
\\AU & 1 & .77 &	$\bf{.88}$ &	$\bf{.88}$ &	.33 &	.69 & $\bf{.88}$ & .69 &	-.69 &	.75 &	.71 &	.42 &	.61 &	$\bf{.89}$ &	$\bf{.85}$\\
\\BE & &  1 &	$\bf{88}$ &	$\bf{.90}$ &	.41 &	.27 &	.80 &	$\bf{94}$ &	-.59 &	$\bf{.92}$ &	$\bf{.83}$ &	$\bf{.85}$ &	.23 &	$\bf{.90}$ &	$\bf{.91}$\\
\\DE & &	& 1 &	$\bf{.90}$ &	.61 &	.35 &	$\bf{.98}$ &	$\bf{.86}$ &	-.65 &	$\bf{.85}$ &	.78 &	.61 &	.52 &	$\bf{.86}$ &	$\bf{.99}$\\
\\DK & & &	& 1 &	.50 &	.58 &	$\bf{.87}$ &	$\bf{.84}$ &	-.80 &	$\bf{.93}$ &	.67 &	.77 &	.58 &	$\bf{.99}$ &	$\bf{.88}$\\
\\ES & & & & &	1 &	-.10 &	.61 &	.34 &	-.38 &	.55 &	.05 &	.36 &	.66 &	.37 &	.64\\
\\FI &	& & & & & 1 &	.42 &	.25 &	-.62 &	.34 &	.27 &	.14 &	.60 &	.64 &	.26\\
\\FR &	& & & & & & 1 &	.79 &	-.71 &	.81 &	.73 &	.52 &	.60 &	.82 &	$\bf{.95}$\\
\\UK & & & & & & & & 1 & -.52 &	.82 &	$\bf{.90}$ &	$\bf{.85}$ &	.12 &	$\bf{.86}$ &	$\bf{.86}$\\
\\GR & & &	& & & & & & 1 &	-.82 &	-.38 &	-.56 &	-.62 &	-.76 &	-.60\\
\\IE &	& & & & & & & & & 1 &	.63 &	$\bf{.85}$ &	.43 &	$\bf{.89}$ &	$\bf{.87}$\\
\\IT &	& & & & & & & & & & 1 &	.59 &	-.05 &	.73 &	.77\\
\\LU &	& & & & & & & & & & & 1 &	.06 &	.77 &	.65\\
\\NL &	& & & & & & & & & & & & 1 &	.50 &	.47\\
\\PT &	& & & & & & & & & & & & &	1 &	$\bf{.84}$\\
\\SE &	& & & & & & & & & & & & & &	1\\
\\\hline

\end{tabular}
\end{center}
\end{footnotesize}
\end{table}

Let us recall that for systems with discrete degrees of freedom, denoted by \textit{s}, the statistical mechanical models are generally defined through the Hamiltonian $H=H(s)$, which is typically a sum of terms, each involving a small number of variables. A useful representation is given by the \textit{factor graph} \cite{pel}. A factor graph is a bipartite graph made of variable nodes $i, j, \ldots $ one for each variable, and function nodes $a, b, \ldots $ one for each term of the Hamiltonian. In the present approach the variable nodes are the macroeconomic indicators and the function nodes are the countries. An edge joins a variable node $i$ and a function node $a$ if and only if $i \in a$, i.e., the variable $s_{i}$ appears in $H_{a}$ - the term of the Hamiltonian associated to $a$. The Hamiltonian can then be written as:

\begin{equation}
H=\sum_{a}H_{a}(s_{a}),\  \hbox{with} \ s_{a}=\left\{s_{i}, i \in a\right\} 
\end{equation}

\par In combinatorial optimization problems \cite{pel}, the Hamiltonian plays the role of a \textit{cost function}. In the low temperature limit $ T \rightarrow \infty $, one is interested by only minimal energy states (ground states) having a non-vanishing probability.
\par Usually, a \textit{cluster} $k$ is defined as a subset of the factor graph such that if a function node belongs to $k$, then all the variable nodes $i \in a $ also belong to $k$ (while the converse needs not to be true, otherwise the only legitimate clusters would be the connected components of the factor graph). Here, this condition will be relaxed by partitioning the function nodes after the criterion if it is connected or not to a certain variable node.
\par Once the correlation matrix is constructed, it is natural to ask for the interpretation of its eigenvalues and eigenvectors. Note that since the matrix is symmetric, the eigenvalues are all real numbers. We will call $\textbf{v}_{a}$ the normalized eigenvector corresponding to eigenvalue $ \lambda_{a}$, with $a = 1, 2,\ldots, M.$. The vector $\textbf{v}_{a}$ is the list of the weights $v_{a, i}$ in this linear combination of the different countries. The variance corresponding to such a combination is thus:

\begin{equation}
\sigma_{a}^{2} = \left\langle \left(\sum_{i=1}^{M} v_{a, i} \hat{d}_{i} \right)^{2}\right\rangle = \sum_{i, j = 1}^{M} v_{a, i} v_{a, j} \hat{C}_{i,j} \equiv {\textbf{v}_{a}}\cdot {\textbf{$\hat{C}$}} {\textbf{v}_{a}}
\end{equation}   

Furthermore, using the fact that different eigenvectors are orthogonal, we obtain a set of uncorrelated random fluctuations $e_{a}$, which are the elements of the system constructed from the weights $v_{a, i}$:

\begin{equation}
e_{a}=\sum_{i=1}^{M} v_{a, i} \hat{d}_{i}, \   \hbox{where} \  \left\langle e_{a} e_{b}\right\rangle = \lambda_{a}\delta_{a, b}
\end{equation}

Conversely, one can think of the initial distances as a linear combination of the uncorrelated factors $E_{a}$:

\begin{equation}
\hat{d}_{i} = \sum_{a=1}^{M} v_{a, i} e_{a}
\end{equation}

In this decomposition, usually called "the principal component analysis", the correlated fluctuations of a set of random variables are decomposed in terms of the fluctuations of underlying uncorrelated factors. In the case of the country clustering, the principal components $E_{a}$ could have an economic interpretation in terms of the macroeconomic indicators.

Since, as generally accepted \cite{risk, plerou}, the largest eigenvectors are the ones carrying the useful information, one can try to define clusters on the basis of the structure of these eigenvectors. Often (but not always), the largest one, $\textbf{v}_{1}$, has comparable and of the same sign components on all countries, and defines the largest cluster, containing all countries. The second one, $\textbf{v}_{2}$, which by construction has to be orthogonal to $\textbf{v}_{1}$, may have some of its components positive, and the others negative. This means that a probable move of the countries around the average (global) fluctuations occurs when some countries over-perform the average, and others under-perform it. Therefore, the sign of the components of $\textbf{v}_{2}$ can be used to group the countries in two families. Each family can then be divided further, using the relative signs of $\textbf{v}_{3}$, $\textbf{v}_{4}$, etc.

\begin{figure}
\includegraphics[height=11cm,width=11cm]{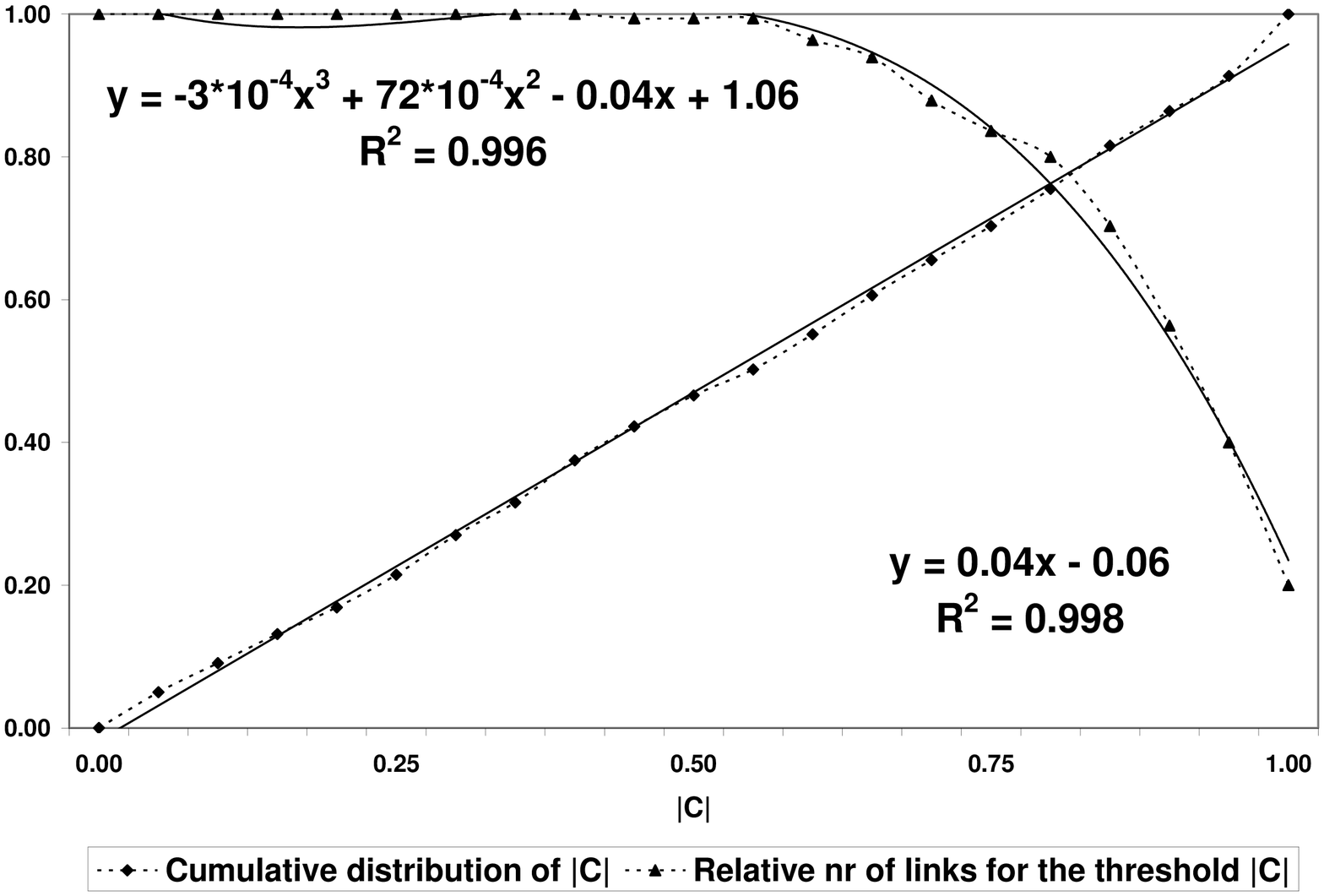}
\caption{\label{GligorFigure1} The cumulative distribution of the correlation coefficients and the relative number of connections versus the $|\hat{C}_{i,j}| \equiv |C|$ (respectively $ C_{thr} \equiv |C|)$ }
\end{figure}

\section{Results}
\subsection{The statistics of the correlation coefficients}

In order to exemplify the MAMPL method, the corresponding steps for x1 = GDP are explicitly shown below. Firstly, the virtual 'AVERAGE' country is introduced in the system. The statistical distances corresponding to the fixed 5 years moving time window are set in increasing order and the minimal length path (MPL) connections to the AVERAGE are established for each country in every time interval (Table I). 

The resulting hierarchy is found to be changing from a time interval to another. Therefore, corresponding correlation matrix is built, this time for the country movements inside the hierarchy (Table II). The above procedure is repeated for each macroeconomic indicator. Thus, the MAMPL method leads us to a set of $M = 11$ correlation matrices, having size $N \times N$, where $N = 15$ is the number of countries under consideration.

Firstly, we analyse the whole set of correlation coefficients. A correlation coefficient $\hat{C}_{i,j}$ will be taken into account as representing a strong connection if and only if $|\hat{C}_{i,j}| > C_{thr}$, where $C_{thr}$ is a certain a priori chosen threshold value. For small values of the $C_{thr}$, all 15 countries have at least one strong connection, i.e. the graph is fully connected. Increasing the $C_{thr}$, the number of the connections decreases. In Fig. 1 the relative number of links (the ratio between the number of actual links and the number of all possible links) is plotted versus the threshold value. One can observe that the data is well fitted by a low order polynomial. In Fig. 1 the cumulative distribution of the correlation coefficients is also plotted (now, the values are the cumulative frequencies and the abscissas are the corresponding correlation coefficients). For comparison, the cumulative uniform distribution is also plotted. The high value of the square of the Pearson product moment correlation coefficient, $R^{2} > 0.99$, indicates a good fit of both distributions.

\begin{figure}
\includegraphics[height=14cm,width=13cm]{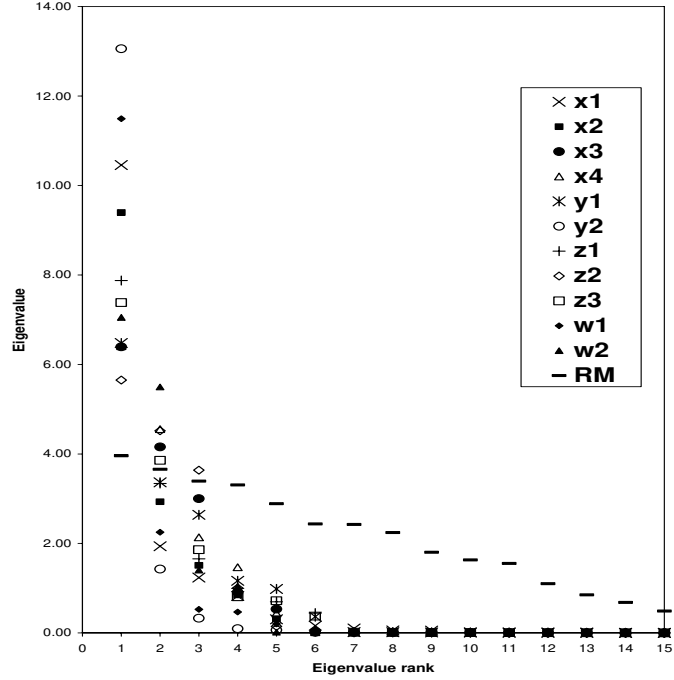}
\caption{\label{GligorFigure2} The eigenvalue spectrum of the correlation matrices between EU-15 country movements with respect to AVERAGE, for each ME indicator (inset). RM: the eigenvalue spectrum of the random matrix.}
\end{figure}

\begin{table}
\caption{The first eigenvector components}
\bigskip
\begin{footnotesize}
\begin{center}
\begin{tabular}{c c c c c c c c c c c c c}

 & GDP & CONS & CAPF	& NEXP	& CPI	& INTR	& LABF	& UNEMP	& GDPH	& GDPC	& GINI
\\\hline 
\\AU & -0.276 & -0.300 &	0.373 &	-0.328 & -0.109 &	-0.274 & 0.239 & 0.305 &	-0.294 &	-0.289 &	-0.261\\
\\BE & -0.287	& -0.325	& 0.357	& 0.189	& 0.003	& -0.271	& 0.308	& 0.229	& -0.351	& -0.259	& -0.371\\
\\DE & -0.296	& -0.304	& 0.257	& -0.371	& -0.334	& -0.274	& -0.343	& 0.299	& -0.284	& -0.261	& -0.122\\
\\DK & -0.303	& -0.097	& 0.281	& 0.111	& -0.003	& -0.276	& -0.293	& -0.250	& -0.161	& -0.287	& -0.131\\
\\ES & -0.167	& -0.325	& 0.356	& -0.171	& -0.260	& -0.276	& 0.331	& -0.271	& 0.244	& -0.275	& 0.360\\
\\FI &	-0.155	& -0.159	& 0.277	& 0.077	& 0.342	& -0.268	& -0.199	& -0.322	& -0.343	& -0.213	& -0.047\\
\\FR &	-0.288	& -0.188	& 0.356	& 0.282	& 0.368	& -0.272	& 0.100	& 0.372	& -0.320	& -0.229	& 0.317\\
\\UK & -0.274	& -0.321	& 0.088	& 0.244	& 0.003	& -0.234	& 0.328	& -0.322	& -0.352	& -0.250	& -0.310\\
\\GR & -0.239	& -0.103	& 0.132	& 0.048	& -0.266	& -0.189	& 0.152	& 0.230	& 0.130	& 0.257	& 0.360\\
\\IE &	-0.290	& -0.325	& 0.274	& 0.351	& 0.300	& -0.276	& -0.163	& -0.322	& 0.068	& -0.282	& 0.188\\
\\IT &	-0.236	& 0.001	& -0.053	& -0.354	& -0.363	& -0.276	& -0.308	& 0.105	& 0.045	& -0.222	& 0.216\\
\\LU &	-0.231	& 0.026	& -0.140	& 0.077	& -0.266	& -0.201	& 0.299	& -0.140	& -0.210	& -0.251	& -0.107\\
\\NL &	-0.165	& -0.325	& 0.059	& 0.056	& 0.110	& -0.274	& 0.151	& -0.194	& -0.207	& -0.272	& -0.345\\
\\PT &	-0.297	& -0.325	& -0.030	& -0.387	& -0.341	& -0.276	& -0.277	& -0.029	& -0.320	& -0.254	& 0.262\\
\\SE &	-0.293	& -0.325	& 0.361	& 0.351	& -0.254	& -0.208	& 0.209	& 0.239	& 0.258	& -0.257	& -0.154\\
\\\hline

\end{tabular}
\end{center}
\end{footnotesize}
\end{table}

\begin{table}

\caption{The second eigenvector components}
\bigskip
\begin{footnotesize}
\begin{center}
\begin{tabular}{c c c c c c c c c c c c c}

 & GDP & CONS & CAPF	& NEXP	& CPI	& INTR	& LABF	& UNEMP	& GDPH	& GDPC	& GINI
\\\hline 
\\AU & 0.014	& -0.155	& 0.043	& -0.030	& -0.285	& -0.079	& 0.393	& 0.268	& -0.204	& -0.078	& 0.121\\
\\BE & -0.236	& -0.042	& -0.124	& 0.279	& -0.179	& -0.074	& -0.026	& -0.060	& -0.086	& 0.224	& 0.051\\
\\DE & 0.013	& -0.141	& 0.204	& -0.110	& -0.162	& -0.046	& 0.009	& 0.273	& 0.174	& 0.295	& 0.339\\
\\DK & 0.052	& 0.335	& -0.315	& -0.433	& 0.387	& 0.003	& -0.238	& 0.335	& 0.276	& -0.099	& -0.397\\
\\ES & 0.247	& -0.033	& 0.146	& -0.094	& -0.234	& -0.032	& -0.040	& -0.197	& -0.192	& -0.232	& -0.083\\
\\FI &	0.404	& 0.427	& -0.306	& -0.423	& -0.164	& -0.114	& 0.359	& -0.054	& 0.006	& -0.424	& -0.385\\
\\FR &	0.079	& 0.142	& 0.146	& 0.012	& -0.149	& -0.086	& -0.256	& -0.012	& 0.194	& 0.268	& 0.190\\
\\UK & -0.309	& 0.039	& -0.420	& -0.191	& 0.085	& 0.314	& -0.110	& -0.061	& -0.011	& 0.092	& 0.103\\
\\GR & 0.238	& 0.332	& 0.266	& -0.356	& 0.241	& -0.605	& -0.399	& -0.358	& 0.340	& 0.283	& -0.083\\
\\IE &	-0.055	& -0.042	& -0.075	& 0.156	& -0.343	& -0.020	& -0.385	& -0.196	& 0.429	& -0.108	& 0.295\\
\\IT &	-0.323	& -0.456	& -0.417	& 0.040	& 0.051	& -0.032	& -0.172	& 0.000	& 0.306	& 0.402	& 0.340\\
\\LU &	-0.306	& 0.560	& -0.090	& -0.423	& -0.309	& 0.471	& -0.113	& 0.424	& 0.392	& 0.199	& 0.300\\
\\NL &	0.576	& -0.033	& -0.264	& -0.372	& -0.448	& -0.079	& -0.355	& 0.381	& -0.352	& -0.186	& 0.109\\
\\PT &	0.007	& -0.033	& -0.438	& 0.052	& -0.094	& -0.032	& 0.129	& 0.443	& 0.126	& -0.323	& -0.241\\
\\SE &	-0.062	& -0.033	& 0.094	& 0.156	& -0.342	& 0.519	& 0.296	& 0.061	& 0.286	& 0.318	& -0.372\\
\\\hline

\end{tabular}
\end{center}
\end{footnotesize}
\end{table}

Nevertheless, performing the $\chi^{2}$ test over the whole set of correlation coefficients we must reject the null hypothesis of the fitting $\left|C\right|$ distribution by the uniform in the confidence interval of 99 \%. Investigating by sight the data set one remarks an anomalous large number of correlation coefficients ($N_{20}=100$) in the range 0.95-1.00, while the mean of the distribution is 57.75 and the standard deviation is $\sigma = 7.45$. According to Chebyshev's theorem \cite{box}, an interval of $\pm 4$ standard deviations ensures that at least 94 \% of the data (of an arbitrary distribution) falls inside this interval. Thus, the last point of the distribution can be treated as an outlier, and, performing the $\chi^{2}$ test for the remainder points we can accept the hypothesis of the same distribution in a confidence interval of over 75 \%. We must note here that the same conclusion is supported by t-Student's test in a confidence interval of 100 \%, the two distributions having \textit{exactly} the same mean. Joining together the results of the statistical tests, we can conclude that the correlation coefficients distribution is a uniform distribution.

\subsection{The bipartite factor graph analysis}

As it has been already shown, the factor graph structure is strongly dependent on the threshold value $C_{thr}$. In order to establish the most appropriate $C_{thr}$, a two tailed t-test of statistical significance is performed over the correlation matrix elements \cite{box}. The null hypothesis (a correlation coefficient of zero) assumes that there is no linear relationship between the two variable sets. In order to test the significance of the correlation coefficients we use the test statistic:

\begin{equation}
t = r \sqrt{\frac{n-2}{1-r^{2}}}
\end{equation}

where $r \equiv \hat{C}_{i,j}$ and $n=2$ is the number of degrees of freedom. The correlation coefficient is considered to be statistically significant if the computed $t$ value is greater than the critical value $t_{C}$ of a t-Student's distribution with a level of significance of $\alpha$. From Eq. (6) one derives:

\begin{equation}
r_{C} = \frac{t_{C}}{\sqrt{t_{C}^{2} + n - 2}}
\end{equation}

Taking $n = 7$ (the number of statistical distances used for computing each correlation coefficient, from the t-Student distribution tables we find the critical value $t_{C}=3.365$ for a reasonable level of significance $\alpha = 0.02$ (or, equivalently, 98 \% confidence interval). From Eq. (7) we get $r_{C} \equiv C_{thr} = 0.83 $ i.e. the null hypothesis can only be rejected for the correlation coefficients greater or at least equal to this value. The significant correlation coefficients are emphasized in bold in Table II.

It is interesting to remark that the two plots from Fig. 1 do intersect at the abscissa 0.83 which is equal to the $r_{C}$ above found. The intersection point seems to correspond to an \textit{optimal} choosing of $C_{thr}$, under the constrain of the competition between link removing and the remainder correlations to be taken into account.

One can easily see that not all 15 countries (function nodes) are connected through the variable node $x_{1}$ (GDP fluctuations), but only 11 of them. Their contributions to the Hamiltonian include the variable $x_{1}$.
\par The above procedure is repeated for each ME variable and leads us to the Hamiltonian (or cost function) having the form: $ H = AUT(x_{1}, x_{2}, x_{3}, x_{4}, y_{2}, z_{1}, z_{2}, z_{3}, w_{1}, w_{2}) + BEL(x_{1}, x_{2}, x_{3}, y_{1}, y_{2}, z_{1}, z_{3}, w_{1}, w_{2}) +
                 DEU(x_{1}, x_{2}, x_{4}, y_{1}, y_{2}, z_{1}, z_{2}, z_{3}, w_{1}, w_{2}) + DNK(x_{1}, x_{3}, x_{4}, y_{2}, z_{1}, z_{2}, w_{1}, w_{2}) +
                 ESP(x_{2}, x_{3}, y_{2}, z_{1}, z_{2}, w_{1}, w_{2}) + FIN(x_{3}, x_{4}, y_{1}, y_{2}, z_{2}, z_{3}, w_{1}, w_{2}) +
                 FRA(x_{1}, x_{3}, x_{4}, y_{1}, y_{2}, z_{2}, z_{3}, w_{1}, w_{2}) + GBR(x_{1}, x_{2}, x_{3}, x_{4}, y_{1}, y_{2}, z_{1}, z_{2}, z_{3}, w_{1}, w_{2}) +
                 GRC(x_{4}, y_{1}, z_{2}, w_{1}, w_{2}) + IRL(x_{1}, x_{2}, x_{3}, x_{4}, y_{1}, y_{2}, z_{2}, w_{1}, w_{2}) +
                  ITA(x_{1}, x_{4}, y_{1}, y_{2}, z_{1}, z_{2}, w_{1}, w_{2}) + LUX(x_{1}, x_{4}, y_{1}, y_{2}, z_{1}, z_{2}, z_{3}, w_{1}, w_{2}) +
                 NLD(x_{2}, x_{4}, y_{2}, z_{2}, w_{1}, w_{2}) + PRT(x_{1}, x_{2}, x_{3}, x_{4}, y_{1}, y_{2}, z_{1}, z_{2}, z_{3}, w_{1}, w_{2}) +
                 SWE(x_{1}, x_{2}, x_{3}, x_{4}, y_{1}, y_{2}, z_{2}, w_{1}, w_{2})$.

\subsection{The correlation matrix analysis}

From the result of the bipartite graph analysis, some countries binary partition in respect to each ME variable can be already seen: a country is connected or not to the respective variable node. Nonetheless, a complete solution to this problem can only be obtained by analyzing the correlation matrix eigensystems. A parallel to similar results from the stock market investigation \cite{risk, plerou} can be also drawn.
\par The eigenvalue spectrum for the empirical correlation matrices is plotted in Fig. 2 for all the ME variables. The results are compared with those of a random uncorrelated matrix (RM), having the same size ($15 \times 15$), constructed by generating random numbers.

\begin{table}

\caption{The EU-15 clustering. The second column displays the eigenvector whose components are used for building the classification scheme. The groups into parentheses are the second-order clusters}
\bigskip
\begin{scriptsize}
\begin{center}
\begin{tabular}{c c c}

 \textbf{INDICATOR} & \textbf{EVC} & \textbf{CLUSTERS}
\\\hline 
\\GDP & $\bf{v_{2}}$ & BEL-GBR-ITA-LUX\\
\\ &  &	AUT-DEU-DNK-FRA-PRT\\
\\ &  &	(ESP-FIN-NLD)\\\hline
\\ Final Consumption & $\bf{v_{2}}$ &	AUT-DEU\\
\\ Expenditure & &	(DNK-FIN-FRA-GRC-LUX)\\
\\\hline
\\ Gross Capital & $\bf{v_{2}}$ &	BEL-DNK-FIN-GBR-PRT\\
\\ Formation & &	ESP-FRA\\
\\\hline
\\ Net Exports & $\bf{v_{1}}$ &	AUT-DEU-ITA-PRT\\
\\ & &	DNK-FRA-GBR-IRL-SWE\\
\\\hline
\\ Consumer Price & $\bf{v_{1}}$ &	DEU-ITA-GRC-LUX\\
\\ Index & &	FIN-FRA-IRL\\
\\\hline
\\ Rate of Interest & $\bf{v_{2}}$ &	GBR-LUX-SWE\\
\\ & &	All the others, except for GRC\\
\\\hline
\\ Labour Force & $\bf{v_{1}}$ &	AUT-BEL-ESP-GBR-LUX\\
\\ & &	DEU-DNK-ITA-PRT\\
\\\hline
\\ Unemployment & $\bf{v_{1}}$ &	AUT-DEU-FRA-GRC-ITA-SWE\\
\\ & &	DNK-ESP-FIN-GBR-IRL-LUX-NLD\\
\\\hline
\\ GDP per hour & $\bf{v_{1}}$ &	DEU-FRA-LUX-PRT\\
\\ worked & &	(ESP-GRC-SWE)\\
\\\hline
\\ GDP per capita & $\bf{v_{2}}$ &	BEL-DEU-FRA-GRC-ITA-LUX-SWE\\
\\ & &	ESP-FIN-IRL-NLD-PRT\\
\\\hline
\\ Gini coefficient & $\bf{v_{1}}$ &	AUT-BEL-DEU-DNK-GBR-LUX-NLD-SWE\\
\\ & &	ESP-FRA-GRC-IRL-ITA-PRT\\
\\\hline

\end{tabular}
\end{center}
\end{scriptsize}
\end{table}

In stock market analysis the largest eigenvalue, often called "market effect", is supposed to describe the collective movement of stock prices, because the corresponding eigenvector components have the same sign and approximately the same size. Looking at the first and second eigenvector components (Tables III and IV) one can easily see that, for the ME correlation matrices, the above interpretation is only partially valid, for $x_{1} \equiv$ GDP, $x_{2} \equiv$  Consumption, $x_{3} \equiv$ Capital Formation, $w_{1} \equiv$ GDP/capita and $y_{2} \equiv$  Interest Rates. The fluctuations of these indicators seem to reflect a global similarity, as a result of the so-called "globalization trend". The same result was also found in \cite{mg} for the first four indicators, by another method, namely measuring the mean statistical distances between countries. The fifth indicator analyzed in \cite{mg} was the Net Exports, for which \textit{no occurrence} of this effect was reported - in perfect agreement with the actual results.

\subsection{Clustering method and results}

The clustering scheme can be next elaborated as follows: firstly, the so-called first order clusters are selected using the bipartite factor graph, i.e. meaning the clusters of countries having at least one connection to the respective variable node. The countries are further partitioned after \textit{the sign} and \textit{the magnitude} of eigenvector components, using Table IV (for $x_{1}, x_{2}, x_{3}, y_{2}$ and $w_{1}$) and Table III (for the others). For several indicators ($x_{1}, x_{2}$ and $z_{3}$) we also selected some groups that can be called second-order clusters, including some countries which are not tied in the factor graph, but have important contributions to the eigenvector structure i.e. large size components. These clusters are written into parentheses in Table V.

Looking at the development indicators ($x_{1}, x_{2}, x_{3}, x_{4}$ and $w_{1}$), we find approximately the same clustering scheme as reported in \cite{mg} but more extended. There is some agreement with the results reported by Chen in \cite{chen} regarding the co-movement between real activity and prices during the period 1992-1997 i.e. the partition of FRA-DEU and ITA into different clusters with respect to the Consumer Price Index fluctuations. Moreover there is agreement with the MST constructed in \cite{hill} for 1996 i.e. the strong connections BEL-DEU-FRA-LUX, IRE-FIN and ESP-PRT with respect to the GDP/capita. 

\section{Concluding remarks}

Here above we have shown that short and noisy macroeconomic time series can be efficiently investigated by moving a constant size time window with a constant step over the time span of interest. The statistical distances between countries, which are calculated using the linear correlations between the datasets for each time interval, can be used for computing the ultrametrical distance from each country to a virtual introduced one, called "Average". This method, called Moving-Average-Minimal-Length-Path, results in a new set of correlation matrices between country distances to their own mean. The new correlation coefficients describe as well as possible the cross-country similarities between the macroeconomic indicator fluctuations around the average common trend.
\par The distribution of the absolute values of the correlation coefficients is the uniform distribution. This can be an effect due to the relative small number of data used for computing them (see Table I), but can be also seen as reflecting the diversity resulted from the large number of particular factors underling the time evolution of each ME indicator. As well as in the biological systems, the existence of some common patterns does not exclude the idiosyncratic diversity.
\par The Bipartite Factor Graph connects in the simplest possible way all the countries by means of corresponding variable nodes assimilated here to the ME indicators. In spite of its simplicity, the method requires an appropriate choosing of the threshold value for the correlation coefficients. One way of evaluating the threshold value can be the t-Student's test of statistical significance, as it has been done in the previous section. We have found the threshold value near 0.83, in a confidence interval of 98 \% of the correlation coefficients statistical significance.
\par The Bipartite Factor Graph leads to a clustering scheme in which \textit{all} the countries are involved (a country can only be tied or not tied to the respective variable). For a reliable clustering scheme, more investigation is required, particularly concerning the tied countries. This investigation was performed in the previous section by analyzing the correlation matrix eigensystems. 
\par As compared with the similar investigation of stock prices clustering, there are some similarities, but also important differences. The Random Matrix Theory could only be partially used here, except for those results valid in the limit of infinite matrices: the finite size effects are much stronger here than in the stock market they are. For finding the so-called noise band \cite{risk}, we had to construct the $N \times N$ ($N = 15$) random matrix having all its rows and columns uncorrelated. Its eigenvalue spectrum was plotted in Fig. 2.
\par The first two eigenvalues (the largest) are far outside the noise band, thus the so called \textit{chance} or \textit{noise correlation} hypothesis can be rejected. Unlike the result obtained for stocks, here the largest eigenvalues does not reflect always a collective mode of the system. The few indicators for which this propriety holds, are the ones more sensitive to the globalization phenomena.
\par Finally, as regards the clustering structure, some overlapping with similar results reported in the economic literature was found. However, the clusters composition is most likely a variable from a time span to another. What is important is the \textit{existence} of the clusters themselves, as this hierarchical structure emerged in a period in which the globalization tendencies were strong and the European common policy was generally oriented to extension and cohesion. In spite of all convergent economic policies, the emergence of the clustering structure seems to be inherent to EU-15 system, as well as it is inherent, perhaps, to any human community.

\end{document}